\documentclass[aps,onecolumn,preprint,pre,superscriptaddress, nofootinbib]{revtex4-1}
 \usepackage{scrextend}
 
\usepackage{verbatim}
 \usepackage{graphicx}
 \usepackage{hyperref}
 \usepackage{mathtools}
 \usepackage{bm}
 \usepackage{lipsum}

\begin{document}
	\title{  Non Rindler    horizons and radiating black holes  }
	\author{Vaibhav Wasnik}
	\email{wasnik@iitgoa.ac.in}
	 \affiliation{Indian Institute of Technology  Goa}
	\begin{abstract}
		In this work we construct metrics corresponding to radiating black holes whose near horizon regions cannot be approximated by Rindler spacetime. We first  construct  infinite parameter coordinate transformations   from Minkowski coordinates, such that an observer using these coordinates to describe spacetime events measures the Minkowski vacuum to be Planckian.   Utilizing these results, we  construct      family of black holes  that  radiate at spatial infinity.  As an illustration we study    a subset of the black hole solutions  that satisfy the null energy condition.
	\end{abstract}

	\maketitle
	\section{Introduction}
The celebrated Hawking result was that Schwarzschild black holes radiate \cite{hawking}.    Inspired by Hawking's result \cite{fulling}, \cite{davies}, \cite{unruh_original}  suggested  that uniformly accelerating observers being restricted to the Rindler wedge of  Minkowski spacetime, would observe the Minkowski vacuum to be thermal.  This is the celebrated Unruh effect.  The   radiation from black holes could  also   be pinpointed to    the Rindler like form of the near horizon metric of the Schwarzschild black holes.  A small textbook way of seeing this is as follows. Let  the metric signature be $(+,-,-,-)$.   Consider the Schwarzschild metric below   
\begin{eqnarray}
ds^2 = (1-\frac{2GM}{r}) dt^2  - \frac{dr^2}{1-\frac{2GM}{r} } - r^2 d\Omega .
\end{eqnarray}
    Define the coordinate $ \hat{r} = r- 2GM$. The  near horizon Schwarzschild metric becomes
\begin{eqnarray}
ds^2 = ( \frac{ \hat{r}}{2GM}) dt^2  - \frac{ 2GM d\hat{r}^2}{\hat{r} } - r^2 d\Omega. 
\end{eqnarray}
If we define $\rho = \sqrt{\hat{r}}$, the above becomes
\begin{eqnarray}
ds^2 &=& ( \frac{  \rho^2 }{2GM}) dt^2  - 8GM d\rho^2 - r^2 d\Omega =  8GM( (    \rho^2  d(\frac{t}{4GM})^2  -  d\rho^2) - r^2 d\Omega \nonumber \\
\end{eqnarray}
The near horizon spacetime metric is hence Rindler like.  Now define $\tau = it$. Then,
\begin{eqnarray}
ds^2 &=& -8GM( (    \rho^2  d(\frac{\tau}{4GM})^2  +  d\rho^2) - r^2 d\Omega. \nonumber \\
\end{eqnarray}
 In order to prevent a conifold singularity at $\rho = 0$, imaginary time $ \tau   $   should be identified with $\tau  + 8\pi GM $. Since, $t$ is the time measured by clocks of an observer at spatial infinity, we are led to the implication that the observer at spatial infinity would measure a temperature $ T = \frac{1}{8\pi GM }$.

 However, consider the metric below that is constructed in this paper
 	\begin{eqnarray}
 ds^2 &=&[1 +  \sum_{n=1, N}   u_n e^{n(t-r)/R}| (r/2GM - 1)|^{-n}]  \nonumber\\
 &&[1 +  \sum_{n'=1, N'}   v_n' e^{-n'(t+r)/R }| ( r/2GM - 1)|^{-n'}][ (1-\frac{2GM}{r})^{} dt^2 \nonumber \\
 &-&    (1-\frac{2GM}{r})^{-1}dr^2] -r^2 d\Omega^2, \nonumber \\
 \label{metric_intro}
 \end{eqnarray}
with   $u_n,v_n$'s are real.   Despite, the metric diverging at $r=2GM$, the spacetime in the neighbourhood of $r=2GM$ is shown in the paper to be non-singular.   In order to examine the causal structure of spacetime implied by the metric, consider radial null curves for which the $\theta, \phi$ are constant and $ds^2=0$, we then see that $\frac{dt}{dr}= \pm (1-\frac{2GM}{r})$, implying light cones close at $r=2GM$, implying a event horizon,  a feature also seen in Schwarzschild black holes.   It is only when all the  $u_n,v_n$'s  are zero that metric has a Schwarzschild form and describes black holes radiating at spatial infinity with temperature $ T = \frac{1}{8\pi GM }$. However when   $u_n,v_n$'s  are not equal to zero, we couldn't use the conifold trick to decipher   whether the corresponding black holes will radiate. In this work  we will explicitly show that   the black holes whose metrics   could be put in the form above also radiate for   $u_n,v_n$'s  that are non zero.

To do this, we  first  introduce a infinite parameter family of coordinate transformations from Minkowski spacetime, such that an observer recording their   spacetime events using these coordinates \footnote{In this work 'observer recording their spacetime events using coordinates $(x,t)$' will imply that the observer records a spacetime event with coordinates $(x,t)$ (such that observers own position in space is at a constant value of $x$ and $t$ is measured by a clock carried by the observer),  which the Minkowski observer records with coordinates $(X,T)$ (such that observers own position in space is at a constant value of $X$ and $T$ is measured by   a clock carried by the observer ).  } will observe the Minkowskian vacuum to contain particles that follow a Planckian spectrum.   Specifically, we show that  observers recording   spacetime events in lightcone coordinates      in the four quadrants of Minkowskian spacetime as shown in Fig.\ref{figure},  related to the Minkowskian lightcone $U, V$ through
	  	\begin{eqnarray}
	  g(bV) &=& e^{bv}, \nonumber\\
	  h(bU) &=& e^{bu}, \nonumber\\
	  \label{eq6}
	  \end{eqnarray}	
	  for $U, V>0$    and 
	  \begin{eqnarray}
	  g(bV) &=& d e^{-b\bar{v}},  \nonumber\\
	  h(bU) &=& de^{-b \bar{u}}, \nonumber\\
	  \label{eq7}
	  \end{eqnarray}	  
	 for $U, V<0$  and  $v,\bar{ v}, u,\bar{ u}\in[-\infty,\infty]$, whichever quadrant   they appear in, would observe the Minkowski vacuum to be Planckian,    if $g, h $ are odd  functions. This is shown in the section II.  We note that if $g(bV) = bV$ and  $h(bU)=bU$, then Eq.\ref{eq6} $\&$ Eq.\ref{eq7}  are the famous Rindler coordinate transformations from Minkowski spacetime. We use these  results of section II to construct  the above metrics Eq.\ref{metric_intro}. The way this is done is the following. First  spacetime is charted with coordinates $(r,t,\theta,\phi)$  with the assumption that  the observer at spatial infinity ( $r\rightarrow \infty$)    lives in Minkowski spacetime and measures distances along direction of increasing $r$ as change in the coordinate $r$ and  records the time on her/his clocks  using       $t$. Next a spherically symmetric time varying mass distribution is assumed to    curve the spacetime  with a event horizon at $r= 2GM$. This mass distribution   is also required to constrain     the relationship between  null coordinates of a  freely falling observer at the event horizon  to the null coordinates of the observer at infinity   through relations mentioned above.  A  class of metrics so constructed  is given by Eq.\ref{metric_intro}.  This is done in section III. In section IV, we consider energy conditions satisfied by energy momentum tensor implied by the mass distribution corresponding to  these metrics. We consider a subset of solutions   and    show   the null energy condition is satisfied.     We end with conclusions.

	\section{  The coordinate transformations}

\begin{figure}[h]
	\centering
\includegraphics[scale=.75]{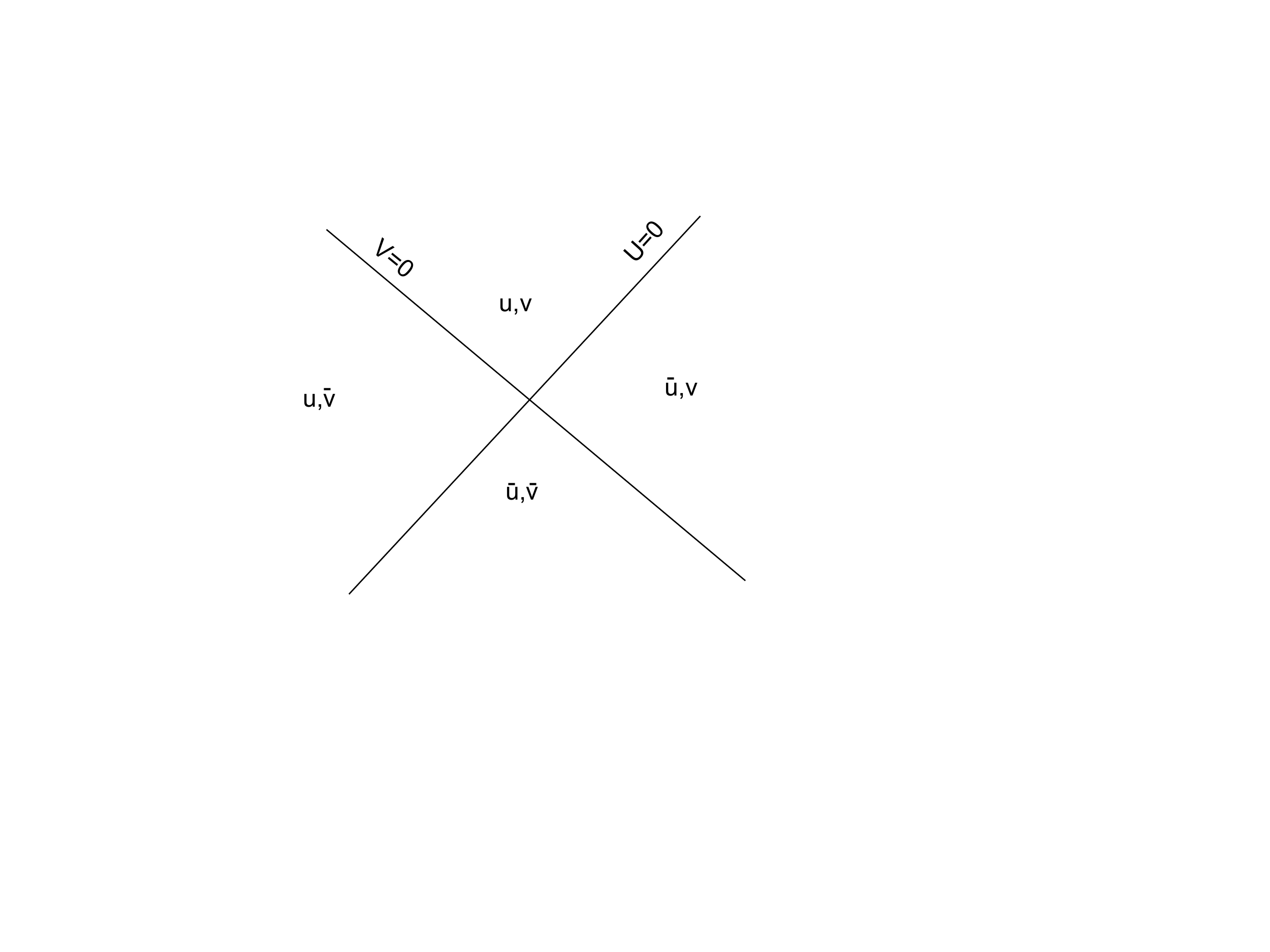}
\caption{ Minkowski spacetime split into four quadrants}
\label{figure}
\end{figure}

Consider the Minkowskian light cone coordinates
  	\begin{eqnarray}
  V &=& T+X,\nonumber\\
  U &=& T-X.\nonumber\\
  \end{eqnarray}
 Consider the four quadrants as shown in Fig.\ref{figure}, with coordinates in each region  
  \begin{eqnarray}
  v, \bar{ v} &=& t+x,\nonumber\\
  u, \bar{ u} &=& t-x,\nonumber\\
  \end{eqnarray}
related to $U,V$ through
	\begin{eqnarray}
g(bV) &=& e^{bv}, \quad V >0  \nonumber\\
h(bU) &=& e^{bu}, \quad U >0\nonumber\\
g(bV) &=& d e^{-b\bar{v}}, \quad V < 0  \nonumber\\
h(bU) &=& de^{-b \bar{u}}, \quad U < 0. \nonumber\\
\label{Eq2}
\end{eqnarray}
 and  $v,\bar{ v}, u,\bar{ u}\in[-\infty,\infty]$, whichever quadrant   they appear in.  Let $d,b$ be real.   $g,h$ are chosen such that,
  \begin{eqnarray}
g(  e^{-i \pi} V) &&= e^{-i(2n+1)\pi} g(V)\nonumber\\
h(e^{-i \pi} U)&&=   e^{-i(2n+1)\pi}  h(U),
\label{eq odd}
  \end{eqnarray}
   with $n$ being a positive integer.   The reason behind the $ e^{-i(2n+1)\pi}$ factors will become clear below.
  

Consider the quadrant in which both $U,V > 0$. We then have 
 \begin{eqnarray}
 ds^2 = dU dV = \frac{ g(bU) h(bV)}{g'(bU) h'(bV)} du dv = f(u,v)du dv,
 \end{eqnarray}
 where $f(u,v) = \frac{ g(bU) h(bV)}{g'(bU) h'(bV)} $  as should be obvious. A Klein Gordon scalar field because of conformal invariance has a action
 
 \begin{eqnarray}
 S &=& \int d^2 x \sqrt{-g} g_{ab} \partial^a \phi \partial^b \phi \nonumber\\
 &=& \int du dv( \partial_{u} \phi \partial_{v} \phi)\nonumber\\
 &=& \int dU dV( \partial_U \phi \partial_V \phi).\nonumber
 \end{eqnarray}
 This implies, any functions of the form $\zeta^U(U)$, $\zeta^V(V)$, $\zeta^u(u)$, $\zeta^v(v)$ are solutions of the Klein Gordon equation.    What is said above can be extended to all quadrants.  This implies, we can talk about positive frequency modes $e^{-i\omega v}$ and $e^{-i\omega \bar{ v}}$, which because of  Eq.\ref{Eq2} could be written as
 \begin{eqnarray}
  \theta(V)e^{-i\omega v} = \theta(V)(g(bV))^{-i\omega/b} &=& \int_0^\infty \frac{dk}{\sqrt{4 \pi k}} [\alpha_{k\omega}^R e^{-ikV} + \beta_{k\omega}^R e^{ikV} ]\nonumber\\
  \label{bogoliubov22}
   \end{eqnarray}
    \begin{eqnarray}
  \theta(-V) e^{-i\omega \bar{ v}}=  \theta(-V)(\frac{g(bV)}{d})^{i\omega/b} &=& \int_0^\infty \frac{dk}{\sqrt{4 \pi k}} [\alpha_{k\omega}^L e^{-ikV} + \beta_{k\omega}^L e^{ikV} ].\nonumber\\
 \label{bogoliubov23}
 \end{eqnarray}
 
 From the bottom equation   we get,
  \begin{eqnarray}
   \theta(V)(e^{-i(2n+1)\pi }\frac{g(bV)}{d})^{i\omega/b} &=& \int_0^\infty \frac{dk}{\sqrt{4 \pi k}} [\alpha_{k\omega}^L e^{ikV} + \beta_{k\omega}^L e^{-ikV} ]\nonumber\\
   \label{relevant_1}
 \end{eqnarray}
  Now when $g(bV)>0 $ for a particular value of $V>0$,  $\ln g(bV)$  is real. Hence we can write $(g(bV))^{-i\omega/b} = e^{-i\omega \ln g(bV)/b }$ and take the complex conjugate of equation   Eq.\ref{bogoliubov22} by changing the sign of the exponent to get 
 \begin{eqnarray}
 \theta(V)(g(bV))^{i\omega/b} &=& \int_0^\infty \frac{dk}{\sqrt{4 \pi k}} [{\alpha_{k\omega}^R}^* e^{ikV} + {\beta_{k\omega}^R}^* e^{-ikV} ],\nonumber\\
 \label{relevant_2}
 \end{eqnarray}
  If however   $g(bV)<0 $ for a particular value of $V$, then $g(-bV)>0 $, so   Eq.\ref{bogoliubov22} can be written as
  \begin{eqnarray}
(-1)^{-i\omega/b}  \theta(V)(g(-bV))^{-i\omega/b} &=& \int_0^\infty \frac{dk}{\sqrt{4 \pi k}} [{\alpha_{k\omega}^R} e^{-ikV} + {\beta_{k\omega}^R} e^{ikV} ],\nonumber\\
  \end{eqnarray}
  respectively. Then we can write $(g(-bV))^{-i\omega/b} = e^{-i\omega \ln g(-bV)/b }$ and we can take the complex conjugate of the above equation by changing the sign of the exponent and we get,
  
   \begin{eqnarray}
  (-1)^{-i\omega/b}  \theta(V)(g(-bV))^{i\omega/b} &=& \int_0^\infty \frac{dk}{\sqrt{4 \pi k}} [{\alpha_{k\omega}^R}^* e^{ikV} + {\beta_{k\omega}^R}^* e^{-ikV} ],\nonumber\\
    \theta(V)(-g(-bV))^{i\omega/b} &=& \int_0^\infty \frac{dk}{\sqrt{4 \pi k}} [{\alpha_{k\omega}^R}^* e^{ikV} + {\beta_{k\omega}^R}^* e^{-ikV} ],\nonumber\\
     \theta(V)(g(bV))^{i\omega/b} &=& \int_0^\infty \frac{dk}{\sqrt{4 \pi k}} [{\alpha_{k\omega}^R}^* e^{ikV} + {\beta_{k\omega}^R}^* e^{-ikV} ],\nonumber\\
  \end{eqnarray}
Hence whether $g(bV)$ is positive or negative we always have
\begin{eqnarray}
\theta(V)(g(bV))^{i\omega/b} &=& \int_0^\infty \frac{dk}{\sqrt{4 \pi k}} [{\alpha_{k\omega}^R}^* e^{ikV} + {\beta_{k\omega}^R}^* e^{-ikV} ],\nonumber\\
\end{eqnarray}
Comparing to Eq.\ref{relevant_1} we get

 \begin{eqnarray}
{\alpha_{k\omega}^R}^* &=& ( e^{i(2n+1)\pi }d)^{i\omega/b} \alpha_{k\omega}^L \nonumber\\
{\beta_{k\omega}^R}^* &=& ( e^{i(2n+1)\pi } d)^{i\omega/b} \beta_{k\omega}^L. \nonumber\\
\label{complex conjugates}
 \end{eqnarray}
 	Upon taking Fourier transforms of Eqs.\ref{bogoliubov22}, \ref{bogoliubov23} we have
 \begin{eqnarray}
 \alpha_{k\omega}^R &=& \sqrt{\frac{k}{\omega}}\int_0^\infty \frac{dV}{2 \pi} (g(bV) )^{-i\omega /b} e^{ikV} \nonumber\\
 \beta_{k\omega}^L &=& \sqrt{\frac{k}{\omega}}\int_{-\infty}^0 \frac{dV}{2 \pi}   (\frac{g(bV)}{d})^{i\omega /b} e^{-ikV}  
 \end{eqnarray}
 Hence
 \begin{eqnarray}
 {	\beta_{k\omega}^L}&=& \sqrt{\frac{k}{\omega}}\int_{-\infty}^0 \frac{dV}{2 \pi}   (\frac{g(bV)}{d} )^{i\omega /b} e^{-ikV}  =  -\sqrt{\frac{k}{\omega}}\int_0^{-\infty}\frac{dV}{2 \pi}   (\frac{g(bV)}{d})^{i\omega /b} e^{-ikV}
 \end{eqnarray}

A question to ask is the following: Does $  (\frac{g(bV)}{d})^{-i\omega /b}$ blow up for any value of $V$?  Let, $V$ be extended to the the whole complex plane then  $g(bV) = r_V e^{i \theta_V}$, where   $r_V \in [0, \infty], \theta_V \in [0,2\pi]$. Now,
\begin{eqnarray}
g(bV)^{-i\omega b } = e^{-i\omega b ln r_V} e^{\theta_V b \omega }= e^{\theta_V b \omega }[ -i\sin [b\omega \ln r_V] +\cos [b\omega \ln r_V] ]
\end{eqnarray}
Since $\theta_V$ is finite and the $\sin, \cos$'s are also finite, $g(bV)^{-i\omega b } $ is finite and hence non-singular over the entire complex plane. Hence,  for a contour $C$ closing in the lower half complex $V$ plane, made up of  the entire real axis and   a semicircle of infinite radius  
 \begin{eqnarray}
 \oint\limits_{C} \frac{dV}{2 \pi}   ( g(bV) )^{i\omega /b} e^{-ikV} = 0
 \end{eqnarray}
 This gives
 \begin{eqnarray}
  {	\beta_{k\omega}^L}&=& \sqrt{\frac{k}{\omega}}\int_{-\infty}^0 \frac{dV}{2 \pi}   (\frac{g(bV)}{d} )^{i\omega /b} e^{-ikV}  =  -\sqrt{\frac{k}{\omega}}\int_0^{-\infty}\frac{dV}{2 \pi}   ( \frac{g(bV)}{d} )^{i\omega /b} e^{-ikV} \nonumber\\
  &=& -\sqrt{\frac{k}{\omega}}\int_0^{\infty}\frac{dV}{2 \pi}   ( \frac{g(bV)}{d} )^{i\omega /b} e^{-ikV} =  -d^{-i \omega /b } {	\alpha_{k\omega}^R }^*\nonumber\\
 \end{eqnarray}
 Using Eq.\ref{complex conjugates}, we have
  
   \begin{eqnarray}
  {	\beta_{k\omega}^R}^*   (e^{i(2n+1)\pi }d)^{-i \omega /b } &=& -d^{-i \omega /b } ( e^{i(2n+1)\pi } d)^{i\omega/b} \alpha_{k\omega}^L 
  \nonumber\\
  {	\beta_{k\omega}^R}^*     &=& -  d^{i \omega /b } ( e^{i(2n+1)\pi })^{i2\omega/b} \alpha_{k\omega}^L 
  \nonumber\\
  \end{eqnarray}
  Now if we assume that $ [d^{i \omega /b } (e^{i(2n+1)\pi } )^{i2\omega/b}  ]^*= d^{-i \omega /b } = e^{-\frac{(2n+1)\pi \omega}{b}}$, then $d$ is constrained to be  $d=e^{ -i(2n+1)\pi}$, and   the following are a   linear combination of positive energy solutions in Minkowski spacetime. 
  \begin{eqnarray}
  G_1(V) &=& \theta(V) e^{-i\omega v}  + \theta(-V) e^{-\frac{(2n+1)\pi \omega}{b}} e^{i\omega \bar{v}} \\
  G_2(V) &=&\theta(V)e^{-\frac{(2n+1)\pi \omega}{b}} e^{i\omega v} +  \theta(-V)  e^{-i\omega \bar{v}} \\
  \label{positive}
  \end{eqnarray}
  
   Now, any field operator can be expressed as 
  \begin{eqnarray}
  \phi(V) &=& \int d\omega[ (\alpha^R_\omega e^{-i\omega v} +  {\alpha^R_\omega} ^\dagger e^{i\omega v}) \theta(V)  \nonumber \\
  &+&(\alpha^L_\omega e^{-i\omega \bar{ v}} + {\alpha^L_\omega} ^\dagger  e^{i\omega \bar{ v}}) \theta(-V)),\\
  &\propto& \int d\omega (  \alpha^R_\omega  - e^{-\frac{(2n+1)\pi \omega}{b}} {\alpha^L_\omega} ^\dagger   )G_1(V) \nonumber \\
  &+& ({a_\omega^L} - e^{-\frac{(2n+1)\pi \omega}{b}}    {a_\omega^R}^\dagger   ) G_2(V)  + h.c.
  \end{eqnarray}
  
  Because $G_{1,2}(V)$ contain Minkowski  modes of    positive frequency  we should have, 
  \begin{eqnarray}
  (  a_\omega^R   - e^{-\frac{(2n+1)\pi \omega}{b}}  {a_\omega^L}^\dagger    |0)\rangle_M = 0
  \end{eqnarray} 
  and
  \begin{eqnarray}
  ( {a_\omega^L}   - e^{-\frac{(2n+1)\pi \omega}{b}} {a_\omega^R}^\dagger      ) |0\rangle_M = 0.
  \end{eqnarray} 
  We could hence, evaluate the occupation number 
  \begin{eqnarray}
  \langle 	{a_\omega^L}^\dagger   	a_\omega^L  \rangle_M = \langle 	{a_\omega^R}^\dagger   	a_\omega^R  \rangle_M = \frac{1}{  e^{2\frac{(2n+1)\pi \omega}{b}}-1},
  \label{Boltzmann}
  \end{eqnarray}

One can easily see that in the derivation above, if we had replaced $v,\bar{ v},V$ with $u,\bar{ u}, U$ respectively, along with $g(bV)$   by $h(bU)$, the expectation value of the corresponding number operators is also Planckian. 
This implies that observer recording spacetime events using corresponding coordinates in respective quadrants in Fig.\ref{figure} would measure a temperature $\frac{ b}{2(2n+1) \pi}$  in a Minkowiskian vacuum if functions $g,h$ obey Eq.\ref{eq odd}  and $d=-1$.

    One can see that we get  led to the same conclusions above by a different route. Let us assume that  as before    $g( e^{-i\pi} bV) = e^{-i(2n+1)\pi} g(   bV) $ below.  Now  consider a function  $G(V)$   made up of positive frequency modes. This function by construction  is analytic in the complex $V$ plane, below the real axis. There is no inconsistency with defining $G(V)   =  g(bV)^{-i\omega/b} $, below the real axis. The reason being that if $V = |V| e^{i\theta}$, $ \theta \in [-
\pi, \pi]$, then there is a branch cut on the negative real axis, as value of $G(V)$ takes two different values at $\theta = \pi $ and $\theta = -\pi$. So, $G(V)   =  g(bV)^{-i\omega/b} $ is not analytic on the whole complex plane.  However, below the real $V$ axis $g(bV)^{-i\omega/b}$ is analytic. This implies that for $V>0$, $G(V) =e^{-i\omega v}  =  g(bV)^{-i\omega/b} $ is a consistent choice for  a function made up of positive frequency Minkowski modes. For $V<0$,

\begin{eqnarray}
G(V)= (g(bV) )^{-i \omega /b } = (- g( e^{-i\pi}  bV) )^{-i \omega /b } =[ e^{-i(2n+1)\pi} (-g(   bV) ) ]^{-i \omega /b } =e^{-\frac{(2n+1)\omega \pi}{b}} e^{ i\omega \bar{v}} \nonumber\\
\end{eqnarray}
   implying
  \begin{eqnarray}
 G_1(V) &=& \theta(V) e^{-i\omega v}  + \theta(-V)   e^{-\frac{(2n+1)\omega\pi}{b}} e^{i\omega \bar{v}}, \\
 \end{eqnarray}
is made up of positive frequency Minkowski modes. One can similarly show that 
 \begin{eqnarray}
G_2(V) &=& \theta(-V) e^{-i\omega \bar{v} }  + \theta(V)   e^{-\frac{(2n+1)\omega\pi}{b}} e^{i\omega v}, \\
\end{eqnarray}
is made up of positive frequency Minkowski modes. Hence, temperature measured by an observer using  the  appropriate  coordinates in the quadrants in Fig.\ref{figure}, is $k_B T  = \frac{ b}{2(2n+1)\pi}$.

  Let us consider a generic case of $g(e^{-i\pi} x) = e^{-i\pi} g(x)$, $h(e^{-i\pi} x) = e^{-i\pi} h(x)$ with,
\begin{eqnarray}
y=g(x)&=& a_1 x + a_3 x^3 +....\nonumber\\
y= h(x) &=& b_1 x + b_3 x^3 +....\nonumber\\
\label{seriesrev}
\end{eqnarray}
Using a series reversion \cite{math_methods_1}, one can then evaluate
\begin{eqnarray}
x = G(y) &=&v_1 y + \frac{v_2}{3} y^3 +....\nonumber\\
x = H(y) &=& u_1 y + \frac{ u_2}{3} y^{3} +....\nonumber\\
\label{nextsection}
\end{eqnarray}
One can similarly start with eqn's above involving the $u,v$'s and can produce the equations with $a,b$'s above using series reversion. We will use the above in the next section.

   One should note that  for $U,V >0$, if for two   functions $J,K$
   \begin{eqnarray}
bV = && J(bv)\nonumber\\
bU = && K(bu),\nonumber\\
   \end{eqnarray}
   then
     \begin{eqnarray}
   dVdU  = && J'(bv)K'(bu)du dv,\nonumber\\
   \end{eqnarray}
   hence the metric in coordinates $u,v$ may not have a time like killing vector. However one should note that the  a time like killing vector is advertised as a requirement so that    that the solution of the wave equation will split up into temporal and spatial parts. This splitting is needed,  so that one could talk about positive/negative frequency modes, which is  a requirement for particle definition under field quantization \cite{de witt}. However note that in two dimensions, the conformal form of the metric above would imply that because  the free scalar field equation is
   \begin{eqnarray}
   \partial_u \partial_v \phi = 0
   \end{eqnarray}
   one would still get   solutions of positive and negative frequencies, with respect to the time coordinates $t$ and hence one could safely talk about particle definitions for an observer using coordinates $(u,v)$ or $(x,t)$.  
 \section{Construction of black hole metrics}
 In this section we will use the results from the previous section to construct metrics corresponding to radiating black holes.
    Consider the Schwarzschild metric
 
 \begin{eqnarray}
 ds^2 = (1-\frac{2GM}{r}) dt^2 - \frac{1}{  1-\frac{2GM}{r}}dr^2 -r^2 d\Omega^2.
 \end{eqnarray}
 In tortoise coordinates
  \begin{eqnarray}
 ds^2 = (1-\frac{2GM}{r}) [dt^2 -  dr_*^2] -r^2 (d\theta^2 + \sin^2 \theta d\phi^2),
 \label{metric}
 \end{eqnarray}
 where 
 
\begin{eqnarray}
r_* &=& r + R \ln|(r/R - 1)|,\nonumber \\
R &=& 2GM.
\end{eqnarray}

The horizon in these coordinates lies at $r_* \rightarrow -\infty$ and spatial infinity at $r_* \rightarrow \infty$. The solution of the Klein Gordon equation in these coordinates has no support behind the horizon. The Klein Gordon action  is
\begin{eqnarray}
S &=& \int d^4 x \sqrt{-g} g^{ab} \partial_a \Phi \partial_b \Phi \nonumber\\
	S &=& \int dr r^2 \sin \theta d\theta d\phi {(1-\frac{2GM}{r}) }  [ \frac{ (\partial_t \Phi)^2 - (\partial_{r^*} \Phi)^2 }{ (1-\frac{2GM}{r})} - \frac{1}{r^2 \sin^2 \theta} (\frac{\partial \Phi }{\partial \phi})^2 -\frac{1}{r^2 } (\frac{\partial \Phi }{\partial \theta })^2 ]\nonumber\\
&=& \int dr r^2 \sin \theta d\theta d\phi [ { (\partial_t \Phi)^2 - (\partial_{r^*} \Phi)^2 }{  } - \frac{(1-\frac{2GM}{r}) }{r^2 \sin^2 \theta} (\frac{\partial \Phi }{\partial \phi})^2 -\frac{(1-\frac{2GM}{r}) }{r^2 } (\frac{\partial \Phi }{\partial \theta })^2 ]\nonumber\\	
\end{eqnarray}
The Klein Gordon equation is hence
\begin{eqnarray}
\partial_a(\sqrt{-g} g^{ab}  \partial_b \Phi )&& = 0\nonumber\\
r^2 \sin \theta \partial_{t}^2 \Phi - \sin \theta \partial_{r^*} r^2 \partial_{r^*} \Phi -r^2 \sin \theta\partial_\phi \frac{(1-\frac{2GM}{r}) }{r^2 \sin^2 \theta} (\frac{\partial \Phi }{\partial \phi})-r^2 \partial_\theta \sin \theta  \frac{(1-\frac{2GM}{r}) }{r^2 } (\frac{\partial \Phi }{\partial \theta })&&=0 \nonumber\\
\partial_{t}^2 \Phi - \frac{1}{r^2}\partial_{r^*} r^2 \partial_{r^*} \Phi -(1-\frac{2GM}{r})[\frac{1}{r^2 \sin^2 \theta }  \partial_\phi  (\frac{\partial \Phi }{\partial \phi})+ \frac{1 }{r^2 \sin \theta } \partial_\theta \sin \theta   (\frac{\partial \Phi }{\partial \theta })] &&=0 \nonumber\\
\label{KG}
\end{eqnarray}
we can see that it would be possible to write  solutions of the above going as $e^{-i\omega t} F_{\omega,l}^{1,2} (r^*)Y^{l,m}(\theta, \phi)$. Here, $F_{\omega,l}^{1,2} (r^*)$ are the two linearly independent solutions of 
\begin{eqnarray}
-\omega^2 F_{\omega,l}^{1,2} (r^*)- \frac{1}{r^2}\partial_{r^*} r^2 \partial_{r^*} F_{\omega,l}^{1,2} (r^*) +(1-\frac{2GM}{r})  \frac{l(l+1)}{r^2}F_{\omega,l}^{1,2} (r^*)&&=0 \nonumber\\
\end{eqnarray}
Hence it would be possible to write the field operator expansion in terms of creation and destruction operators for field quanta as

\begin{eqnarray}
\phi(r^*,\phi, \theta, t) &=& \sum_{\omega,l,m} [a_\omega^{l,m} e^{-i\omega t} F^1_{l,\omega} (r^*,\theta, \phi) + b_\omega^{l,m}  e^{-i\omega t} F^2_{l,\omega} (r^*,\theta, \phi) +  h.c.] \nonumber\\
&&Y^{l,m}(\theta, \phi)\nonumber\\
\end{eqnarray}

Close to the horizon $r=2GM$ implies the angular derivatives contribution in Eq.\ref{KG}, become zero and $\frac{dr}{dr^*} = 0$. Hence,   the Klein Gordon equation   becomes

\begin{eqnarray}
(\partial_{t}^2 -\partial_{r^*}^2) \phi = 0
\end{eqnarray}
At spatial infinity, the metric Eq.\ref{metric} is also Minkowskian. This situation is equivalent to a free particle coming from either asymptotes encountering a potential barrier, implying two independent solutions that asymptote as

\begin{eqnarray}
	F_{\omega,l}^{1} (r^*) &=& e^{i \omega r^*} + A(l,\omega) e^{-i\omega r^*}\quad r^* \rightarrow -\infty \nonumber\\
	&=& e^{i \omega r^*} \quad r^* \rightarrow \infty \nonumber\\
\end{eqnarray}
and 
\begin{eqnarray}
F_{\omega,l}^{2} (r^*) &=& e^{-i \omega r^*} + B(l,\omega) e^{i\omega r^*}\quad r^* \rightarrow \infty \nonumber\\
&=& e^{-i \omega r^*} \quad r^* \rightarrow -\infty \nonumber\\
\end{eqnarray}

Hence close to the horizon we can expand
\begin{eqnarray}
\phi(r^*\rightarrow -\infty,\phi, \theta, t) &=& \sum_{\omega ,l,m }[   a_\omega^{l,m}  [ e^{i \omega r^*} + A(l,\omega) e^{-i\omega r^*}] e^{-i\omega t} + b_\omega^{l,m}   e^{-i \omega (r^*+t) } + h.c. ] Y^{l,m}(\theta, \phi) \nonumber\\
 &=& \sum_{\omega ,l,m }[   a_\omega^{l,m}   e^{i \omega (r^*-t)} + (b_\omega^{l,m}  + a_\omega^{l,m}  A(l,\omega)) e^{-i\omega (r^*+t)}] + h.c. ] Y^{l,m}(\theta, \phi) \nonumber\\
 \label{rindler expansion}
\end{eqnarray}

and at spatial infinity,  
\begin{eqnarray}
\phi(r^*\rightarrow \infty,\phi, \theta, t) &=&   \sum_{\omega ,l,m }[   b_\omega^{l,m}  [ e^{-i \omega r^*} + B(l,\omega) e^{i\omega r^*}] e^{-i\omega t} + a_\omega^{l,m}   e^{i \omega (r^*-t) }+h.c. ] Y^{l,m}(\theta, \phi) \nonumber\\
&=&   \sum_{\omega ,l,m }[  b_\omega^{l,m}   e^{-i \omega (r^*+t)} +(a_\omega^{l,m}  + b_\omega^{l,m}  B(l,\omega) )e^{i\omega (r^*-t)}  + h.c. ] Y^{l,m}(\theta, \phi) \nonumber\\
 \label{minkowski expansion}
\end{eqnarray}
This implies that the $a_\omega^{l,m} , b_\omega^{l,m} $ are particle destruction operators    both at the horizon as well as spatial infinity. 

  Close to and outside the horizon, the metric ignoring the $r^2 d\Omega^2$   can be written  as 
  \begin{eqnarray}
ds^2 &=& (1-\frac{R}{r}) [dt^2 -  dr_*^2]   \\
&\sim& e^{ (r_*-r)/R}[dt^2 -  dr_*^2]   \\
&\sim&   e^{ (v-u)/2R-1}[dudv]   \nonumber \\
&\sim & dUdV, 
\end{eqnarray}
where $   U = -R e^{-u/2R}$, $V  =R e^{v/2R}$, are the null   coordinates of the freely falling observer. The $u$,$v$ coordinates hence are of the form of   Rindler coordinates. We could similarly represent the coordinates close to and behind the horizon  as  $  U = R e^{u/2R}$, $ V  =-R e^{-v/2R}$ .  This implies that the expectation values of the number operators  ${a_\omega^{l,m} }^\dagger a_\omega^{l,m} $ and ${b_\omega^{l,m} }^\dagger b_\omega^{l,m} $ in Eq.\ref{rindler expansion}  in the Minkowskian vacuum would have a Planckian form, which will then also imply that the spectrum of particles at spatial infinity is also Planckian.  The above reasoning is inspired from  \cite{jacobsson}, \cite{de witt}, \cite{candelas}.

Now consider the following scenario. What if instead of the metric Eq.\ref{metric}, we had the following metric

 \begin{eqnarray}
ds^2 = \Pi(r,t) [dt^2 -  dr_*^2] -r^2 (d\theta^2 + \sin^2 \theta d\phi^2),
\label{modified metric}
\end{eqnarray}
where $\Pi(r,t)$ is a smooth function of its parameters. The Klein Gordon action is now

\begin{eqnarray}
S &=& \int d^4 x \sqrt{-g} g^{ab} \partial_a \Phi \partial_b \Phi \nonumber\\
S &=& \int dr r^2 \sin \theta d\theta d\phi {\Pi(r,t)}  [ \frac{ (\partial_t \Phi)^2 - (\partial_{r^*} \Phi)^2 }{ \Pi(r,t)} - \frac{1}{r^2 \sin^2 \theta} (\frac{\partial \Phi }{\partial \phi})^2 -\frac{1}{r^2 } (\frac{\partial \Phi }{\partial \theta })^2 ]\nonumber\\
&=& \int dr r^2 \sin \theta d\theta d\phi [ { (\partial_t \Phi)^2 - (\partial_{r^*} \Phi)^2 }{  } - \frac{ \Pi(r,t) }{r^2 \sin^2 \theta} (\frac{\partial \Phi }{\partial \phi})^2 -\frac{ \Pi(r,t) }{r^2 } (\frac{\partial \Phi }{\partial \theta })^2 ]\nonumber\\	
\end{eqnarray}

The Klein Gordon equation is hence
\begin{eqnarray}
\partial_a(\sqrt{-g} g^{ab}  \partial_b \Phi )&& = 0\nonumber\\
\partial_{t}^2 \Phi - \frac{1}{r^2}\partial_{r^*} r^2 \partial_{r^*} \Phi -\Pi(r,t)[\frac{1}{r^2 \sin^2 \theta }  \partial_\phi  (\frac{\partial \Phi }{\partial \phi})+ \frac{1 }{r^2 \sin \theta } \partial_\theta \sin \theta   (\frac{\partial \Phi }{\partial \theta })] &&=0 \nonumber\\
\end{eqnarray}
 If we now try solutions of the form $e^{-i\omega t} F_{\omega,l}^{1,2} (r^*)Y^{l,m}(\theta, \phi)$, then
 \begin{eqnarray}
 -\omega^2 F_{\omega,l}^{1,2} (r^*)- \frac{1}{r^2}\partial_{r^*} r^2 \partial_{r^*} F_{\omega,l}^{1,2} (r^*) + \Pi(r,t)  \frac{l(l+1)}{r^2}F_{\omega,l}^{1,2} (r^*)&&=0 \nonumber\\
 \end{eqnarray}
 We now see that  we could only get a solution for $\Pi(r,t) $ not being independent of $t$, if $l=0$.  We should note that the  $l=0$ restriction only arises if one is considering time dependent metrics which have the generic form  Eq.\ref{modified metric}  and are interested in solutions of Klein Gordon    equation of the form $e^{-i\omega t} F_{\omega,l}^{1,2} (r^*)Y^{l,m}(\theta, \phi)$. If we instead had a metric of the form 
 
 \begin{eqnarray}
 ds^2 = \Pi_t(r,t) dt^2 -   \Pi_r(r,t)dr_*^2 -r^2 (d\theta^2 + \sin^2 \theta d\phi^2),
 \label{above_metric}
 \end{eqnarray}
 with $ \Pi_t(r,t)$,  $ \Pi_r(r,t) $ being two different functions, then it would not be possible to get a solution of the   Klein Gordon equation of the form  $e^{-i\omega t} F_{\omega,l}^{1,2} (r^*)Y^{l,m}(\theta, \phi)$ for any values of $l,m$.  Hence only wave modes of the form $e^{-i\omega t} F_{\omega, l=0}^{1,2}$, will correspond to particles after field quantization  in a background given by metric of the form Eq.\ref{modified metric}. Now if the near horizon form of $\Pi(r,t)$ is such that we could map the lightlike coordinates $u=t+r$, $v=t-r$, to free falling $U,V$ as in  Eq .\ref{Eq2}, we would get thermal radiation   with quantum number $l=0$ at spatial infinity. 
 
 We use these arguments  to construct black hole metrics that correspond to radiating black holes below. 
 
In light of the results from the previous section, we argue that if  the relationships between the freely falling coordinates and $u,v$ coordinates describing the region outside the black hole horizon    were   $   U = -H(e^{-u/2R})$, $V  =G( e^{v/2R})$,  with $H,G$ being odd functions as in    Eq.\ref{nextsection}, an observer  at spatial infinity would measure a temperature. By considering   Eq. 10  that gives the transformation between Minkowski coordinates to the   null coordinates $(u,v)$,  we   derived   the temperature measured by an observer using these coordinates to be   $\frac{ b}{2(2n+1)\pi}$, as was seen from  Eq.\ref{Boltzmann}. Also since  the corresponding $n = 0$, for the choice of coordinate transformations relating the free falling coordinates at the horizon to the coordinates of an observer at infinity, the temperature measured at infinity will go as $\frac{1}{4R\pi}= \frac{1}{8GM\pi}$. 


\begin{eqnarray}
ds^2 &=& dVdU\\
 &\sim &  e^{ (v-u)/2R}H'(e^{-u/2R})G'( e^{v/2R})[dudv].  \nonumber \\
\end{eqnarray}

  	We claim that the following metric
  	\begin{eqnarray}
  	ds^2 &=&[1 +  \sum_{n=1, N}   u_n e^{n(t-r)/r}| (r/R - 1)|^{-n}]  \nonumber\\
  &&[1 +  \sum_{n'=1, N'}   v_n' e^{-n'(t+r)/R }| (r/R - 1)|^{-n'}][ (1-\frac{R}{r})^{} dt^2 \nonumber \\
  &-&    (1-\frac{R}{r})^{-1}dr^2] -r^2 d\Omega^2, \nonumber \\
  \label{actual_metric}
  	\end{eqnarray}
  	will reduce to the above form close to the horizon. To see this, note that because
  	
  	\begin{eqnarray}
  	r_* &=& r + R \ln|(r/R - 1)|,\nonumber \\
  	\end{eqnarray}
  	\begin{eqnarray}
  e^{v/R}	=e^{(t+ r_* )/R}&=& e^{(t+r)/R} |(r/R - 1)|,\\
e^{u/R}	=  	e^{(t- r_* )/R}&=& e^{(t-r)/R} |(r/R - 1)|^{-1}.\nonumber \\
  	\end{eqnarray}
     	
  	This implies, that close to the horizon we have
  	\begin{eqnarray}
  	ds^2 &=&[1 +  \sum_{n=1, N}   u_n e^{n(t-r)/R}| (r/R - 1)|^{-n}]  \nonumber\\
  	&&[1 +  \sum_{n=1, N'}   v_n' e^{-n'(t+r)/R}| (r/R - 1)|^{-n'}][ (1-\frac{R}{r})^{} dt^2 \nonumber \\
  	&-&    (1-\frac{R}{r})^{-1}dr^2 ]-r^2 d\Omega^2, \nonumber \\
  	&\sim &   e^{ (v-u)/2R}[    1 +   \sum_{n=1,N} u_n e^{n u/R} ] [    1 +   \sum_{n'=1,N'} v_n' e^{-n' v/R} ] [dudv] - r^2 d\Omega^2 \nonumber \\
  	&\sim&  e^{ (v-u)/2R}H'(e^{u/2R})G'( e^{-v/2R})[dudv]  - r^2 d\Omega^2 \nonumber \\
  	\end{eqnarray}

At spatial infinity, the above metric by construction becomes the inertial
\begin{eqnarray}
ds^2 &\sim &  [  dt^2 -    dr^2 -r^2 d\Omega^2].\nonumber \\
\end{eqnarray}

 We note that because the non-angular part of  Eq.\ref{actual_metric} close to the horizon is given  $ \sim e^{ (v-u)/2R}H'(e^{u/2R})G'( e^{-v/2R})[dudv]$, with the co-ordinates $u,v$ related to that of flat spacetime $U,V$, as $   U = -H(e^{-u/2R})$, $V  =G( e^{v/2R})$, implies that the near horizon geometry is not singular, despite the fact that the metric blows up at the horizon. Another way to see this is that the metric ignoring the angular parts is basically the Schwarzchild metric multiplied by $[1 +  \sum_{n=1, N}   u_n e^{n(t-r)/R}| (r/R - 1)|^{-n}]  
 [1 +  \sum_{n=1, N'}   v_n' e^{-n'(t+r)/R}| (r/R - 1)|^{-n'}] $, which equals $H'(e^{u/2R})G'( e^{-v/2R})$ globally and not just at the horizon. So this enhancement to the Schwarzchild metric cannot change the non-singular properties of the horizon.
 
 We also note that the discussion from Eq.\ref{modified metric} to above Eq.\ref{above_metric}, still holds even if the $\Pi(r,t)$ diverges at any point in spacetime, as the discussion is about how $\Pi(r,t)$ does not appear in any discussion of $l=0$ wave modes. Hence even though the   the metric Eq.\ref{actual_metric} diverges at the horizon, black holes corresponding to the metric  radiate. 
 
 Because the metric is time dependent and can blow up for $t\rightarrow \infty$ and $t\rightarrow -\infty$ we   assume that the mass distribution  corresponding to the metric above is produced during a physical process (that we do not comment on) at $t=0$ and  this time varying mass distribution later evolves into something else and does not correspond to the metric above   $t >T$ where $T$ is some finite time.  Radiation will be observed  for $t \in [0,T]$.  This statement is consistent and is similar to the statement that   a  body would be bombarded with thermal radiation for the period of time it is moving with a constant acceleration, despite other instances on  its worldline where the body may be moving with a constant velocity. 
 
 \section{Energy Conditions}
 We have the metric
 	\begin{eqnarray}
 ds^2 &=&[1 +  \sum_{n=1, N}   u_n e^{n(t-r)/R}| (r/R- 1)|^{-n}]  \nonumber\\
 &&[1 +  \sum_{n'=1, N'}   v_n' e^{-n'(t+r)/R }| (r/R - 1)|^{-n'}][ (1-\frac{R}{r})^{} dt^2 \nonumber \\
 &-&    (1-\frac{R}{r})^{-1}dr^2] -r^2 d\Omega^2. \nonumber \\
 \end{eqnarray}
   The constraints on physicality of the solution above are imposed by the energy conditions. To evaluate these conditions we need to evaluate the energy momentum tensor corresponding to the above metric. To reduce the clutter of calculations because of so many terms in the sum above, let us for illustration consider the case where all the $v_n = 0$ and only one of $u_n$'s which we label as $a$ is non-zero, which corresponds to $n=2p$, with $p$ being a positive integer. We hence have the metric,
 

 \begin{equation}
\left(
\begin{array}{cccc}
-\frac{a e^{\frac{2 p (t-r)}{R}} \left(1-\frac{r}{R}\right)^{-2 p}+1}{1-\frac{R}{r}} & 0 & 0 & 0 \\
0 & -r^2 & 0 & 0 \\
0 & 0 & -r^2 \sin ^2(\theta ) & 0 \\
0 & 0 & 0 & \left(a e^{\frac{2 p (t-r)}{R}} \left(1-\frac{r}{R}\right)^{-2 p}+1\right) \left(1-\frac{R}{r}\right) \\
\end{array}
\right)
\label{matrix}
 \end{equation}
 Using Mathematica we get the following expressions for  non-zero components of energy momentum tensor  
 
 \begin{eqnarray}
8\pi G T_{rr}  
&=&
\frac{a e^{\frac{2 p (t-r)}{R}} \left(\frac{2 p}{R \left(a e^{\frac{2 p (t-r)}{R}}+\left(1-\frac{r}{R}\right)^{2 p}\right)}+\frac{\left(1-\frac{r}{R}\right)^{-2 p}}{r}\right)}{R-r}\nonumber\\
8\pi G T_{tt}  
&=&
-\frac{a (R-r) \left(1-\frac{r}{R}\right)^{-2 p} e^{\frac{2 p (t-r)}{R}} \left(R \left(a e^{\frac{2 p (t-r)}{R}}+\left(1-\frac{r}{R}\right)^{2 p}\right)-2 p r \left(1-\frac{r}{R}\right)^{2 p}\right)}{r^3 R \left(a e^{\frac{2 p (t-r)}{R}}+\left(1-\frac{r}{R}\right)^{2 p}\right)}\nonumber\\
8\pi G T_{rt} &&= 8\pi G T_{tr}  = 
\frac{2 a p e^{\frac{2 p t}{R}}}{a r R e^{\frac{2 p t}{R}}+r R e^{\frac{2 p r}{R}} \left(-\frac{r-R}{R}\right)^{2 p}}
 \end{eqnarray}
 Now a radial null vector given the metric Eq.\ref{matrix}, is any multiple of $n^\mu= (1-\frac{R}{r},0,0,1)$ (our coordinates are $(r,\theta, \phi, t)$). Then, 
 \begin{eqnarray}
 T_{\nu \mu }n^\nu n^\mu &=&0
 \end{eqnarray}
 We hence see the null energy  condition is satisfied.  
 \section{Conclusion}
 Some alternatives to Rindler  spacetimes such that observers in the spacetime observe a Planckian spectrum were constructed in   \cite{sanchez1}. It was shown in \cite{sanchez2}, that the constraint of global thermal equilibrium, which translates into vanishing of the expectation value of the momentum operator in such spacetimes, however singles out Rindler observers as ones which observe the Minkowski vacuum at thermal equilibrium. However note that global requirement can be relaxed when we are only concerned with behaviours of black hole metrics near the horizon. The fact that particles are emitted by the black hole radially outwards at spatial infinity anyway imply        a non zero expectation value for the momentum operator. We can hence consider non Rindler like   form of metric close to the black hole horizon to comment about the thermal Planckian spectrum of radiation emitted by the black hole as we have   done in this paper.    In the section above we considered one family of solutions by considering only one of the $u_n$'s to be non-zero.  It would be of interest to classify all other   acceptable solutions implied by the metric in  Eq.\ref{actual_metric}.  We note that the temperature for the black holes constructed   is not equal to $T=\kappa/2\pi$, where $\kappa$ is the surface gravity, as the definition of surface gravity needs that the event horizon is a Killing horizon for a Killing vector $\chi^\mu$, with $\kappa^2 = -\frac{1}{2}  \nabla^\mu \chi^\nu  \nabla_\mu \chi_\nu$. However,  only a vector proportional to     $\partial_t$    becomes null at the event horizon located at $r=2GM$ for the metric in Eq.\ref{actual_metric}, but the metric  being non stationary does not have $\partial_t$ as a Killing vector. This implies the event horizon is not a Killing horizon and hence  the traditional definition of surface gravity at the horizon  $\kappa^2 = -\frac{1}{2}  \nabla^\mu \chi^\nu  \nabla_\mu \chi_\nu$    fails for this metric. Hence a     framing of  the zeroth and third law of black hole mechanics is not possible. Since the stress energy tensor is not zero outside the event horizon, one could not construct a law that relates the mass of the black hole to the surface area of the event horizon as these definitions implicity assume that all the mass/energy is located behind the event horizon.  The first law also involves the surface gravity as a proportionality constant between the surface area of the event horizon and the mass of the black hole, and the non existence of a definition of surface gravity further adds to issues in constructing this law for the black hole solution described in the manuscript. The solution however tends to the Schwarzchild metric at $t\rightarrow \infty$. One could hence understand the solutions constructed in the manuscript to describe radiating black holes that do not possess a charge or angular momentum  but eat   up matter before equilibriating to a  Schwarzchild black hole.  As we have mentioned above,   the metric Eq.\ref{matrix} would correspond to   a mass distribution produced during a physical process, that we do not comment on, at $t=0$ and  this time varying mass distribution later evolves into something else and does not correspond to the metric   for   $t >T$ where $T$ is some finite time.  Understanding    physical processes that lead to metrics constructed in the paper   needs further research.

     \section{Conflict of Interests}
     The authors declare no conflict of interest.

	\end{document}